\documentclass[aps,prd,twocolumn,groupedaddress]{revtex4-1}
\usepackage{graphicx}
\include{amssym}

\begin{document}
\title{Ratios of the light quark masses: Cubic curve vs ellipse}
\author{A.\,A. Osipov\footnote{Email address: osipov@nu.jinr.ru}}
\affiliation{Joint Institute for Nuclear Research, Bogoliubov Laboratory of Theoretical Physics, 141980 Dubna, Russia}

\begin{abstract}
The bounds on the light quark masses are obtained by fitting the squares of pseudoscalar meson masses $m_\pi^2$ and $m_K^2$  to second order in $1/N_c$ expansion. The result is an algebraic cubic curve whose coefficients are the known Weinberg values for the quark mass ratios $m_u/m_d$ and $m_s/m_d$. Additional restrictions arise when using the ratio $m_s/m_{ud}=27.23(10)$ quoted by FLAG for lattice simulations with four quark flavors. This provides a tight constraint on the ratio $m_u/m_d = 0.455(8)$. 
 \end{abstract}

\maketitle

\vspace{0.5cm}


\section{Introduction}
The values of the $u$, $d$ and $s$ light quark masses are the fundamental parameters, which are responsible for the explicit $SU(3)_L\times SU(3)_R$ chiral symmetry breaking in the QCD Lagrangian. Their accurate determination is important for both phenomenological and theoretical applications \cite{Gasser:82}. One of the main sources for obtaining information about the magnitude of the light quark masses is the octet of pseudoscalar mesons, i.e., pions, kaons, and eta.  This is due to the special role it plays in particle physics. Being Goldstone modes of spontaneously broken chiral $SU(3)_L\times SU(3)_R$ symmetry, the members of the octet acquire mass only due to non-zero values of the light quark masses. The celebrated Gell-Mann-Oakes-Renner relation \cite{GOR:68} predicts that the squared mass of a Nambu-Goldstone mode, $\mu^2$, is directly proportional to the sum of the masses of the quark and antiquark which compose it, up to electromagnetic and higher-order mass corrections. In particular, for the masses of the $\pi^\pm$, $\pi^0$, $K^\pm$ and $K^0,{\bar K}^0$ mesons we have 
\begin{equation}
\label{GOR}
\bar\mu_{ij}^2=B_0(m_i+m_j).
\end{equation}
Here and hereafter the bar over the mass symbol $\mu$ indicates that electromagnetic corrections in the expressions for the masses of charged states are neglected. The indices $(i, j)$ correspond to the quark content of a particular meson state: $(u,d)\to \pi^\pm, \pi^0$, $(u,s)\to K^\pm$, $(d,s)\to K^0,{\bar K}^0$. The dimensional parameter $B_0$ is associated with the quark condensate $B_0=-\langle \bar qq\rangle /F^2$, where $F$ is the pion decay constant (in the chiral limit). 

To this current-algebra result one can apply Dashen's theorem \cite{Dashen:69}, according to which in the chiral limit virtual photons yield the same electromagnetic correction to the observed masses of charged pions $m_{\pi^\pm}$ and kaons $m_{K^\pm}$, i.e., in the formulas $m^2_{\pi^+}=\bar\mu^2_{\pi^+}+\Delta^2_{\mathrm{em}}$, $m^2_{K^+}=\bar\mu^2_{K^+}+\tilde\Delta^2_{\mathrm{em}}$ one should set $\Delta^2_{\mathrm{em}}=\tilde \Delta^2_{\mathrm{em}}$. If we additionally neglect corrections of order $(m_d-m_u)^2$, then
$\Delta^2_{\mathrm{em}}=m^2_{\pi^+}-m^2_{\pi^0}$. As a consequence, one arrives to Weinberg's values \cite{Weinberg:77} of the quark mass ratios
\begin{eqnarray}
\label{Wv}
x_W&=& \left(\frac{m_u}{m_d}\right)_W=\frac{\bar\mu^2_{K^+}-\bar\mu^2_{K^0}+\bar\mu^2_{\pi^+}}{\bar\mu^2_{K^0}-\bar\mu^2_{K^+}+\bar\mu^2_{\pi^+}} \nonumber \\
&\stackrel{\mathrm{LO}}{=}&\frac{m^2_{K^+}-m^2_{K^0}+2m^2_{\pi^0}-m^2_{\pi^+}}{m^2_{K^0}-m^2_{K^+}+m^2_{\pi^+}}=0.56,  \\
y_W&=& \left(\frac{m_s}{m_d}\right)_W=\frac{\bar\mu^2_{K^+}+\bar\mu^2_{K^0}-\bar\mu^2_{\pi^+}}{\bar\mu^2_{K^0}-\bar\mu^2_{K^+}+\bar\mu^2_{\pi^+}}\nonumber \\
&\stackrel{\mathrm{LO}}{=}&\frac{m^2_{K^+}+m^2_{K^0}-m^2_{\pi^+}}{m^2_{K^0}-m^2_{K^+}+m^2_{\pi^+}}=20.18.
\end{eqnarray}
It also follows that the other two quark mass ratios are 
\begin{eqnarray}
S&\equiv&\frac{m_s}{m_{ud}}\stackrel{\mathrm{LO}}{=}26, \\
R&\equiv&\frac{m_s-m_{ud}}{m_d-m_u}\stackrel{\mathrm{LO}}{=}44, 
\end{eqnarray}
where $m_{ud}$ is the isospin-averaged quark mass $m_{ud}=(m_u+m_d)/2$.

A systematic study of the next-to-leading order (NLO) correction to Weinberg's values in chiral perturbation theory ($\chi$PT), made by Gasser and Leutwyler (GL) \cite{GL:85}, shows that this step constrains the ratios $m_u/m_d$ and $m_s/m_d$ to an ellipse. The ellipse is due to a low-energy theorem stating that there is a certain double ratio of the squared meson masses in which NLO-corrections are negligible ($\sim (m_d-m_u)^2/m_s^2$) and can therefore be discarded  
\begin{equation}
\label{GL}
\frac{m_K^2(m_K^2-m_\pi^2)}{m_\pi^2(\bar m_{K^0}^2- \bar m_{K^+}^2)}
\stackrel{\mathrm{NLO}}{=}\frac{m_s^2-m_{ud}^2}{m_d^2-m_u^2}\equiv Q^2_{\mathrm{GL}}.
\end{equation}
Here $\bar m_{K^0}$, $\bar m_{K^+}$ denote the mass of the neutral and charged kaons in QCD, while $m_\pi$, $m_K$ represent the mass of the pions and kaons in the isospin limit, respectively. The dispersive analysis of $\eta\to 3\pi$ decays \cite{Colangelo:17,Colangelo:18} allows to establish a constraint on the value of the semi-major axis of the ellipse $Q_{\mathrm{GL}}=22.1(7)$. Combined with the known results from lattice calculations for $S=27.30(34)$ with four quark flavors $N_f=2+1+1$ \cite{FLAG:17}, this essentially limits the range of allowed quark mass ratios: $m_u/m_d=0.44(3)$, $R=34.2(2.2)$ \cite{Colangelo:18}.

It should be noted that have been many attempts over the years to extract the ratio $Q_{\mathrm{GL}}$ from $\eta\to 3\pi$ decays. Different methods were applied (see, for example, Table 9 in \cite{Colangelo:18}). Currently, there is some discrepancy between the result accounting for the NNLO corrections in $\chi$PT and those obtained by dispersion calculations. Namely, the NNLO corrections in $\chi$PT (at first order in isospin breaking) lead to somewhat larger effect than that implied by the dispersive estimates. This is reflected in the corresponding result for $Q_{\mathrm{GL}}=23.2$ \cite{Bijnens:07}, compared with $Q_{\mathrm{GL}}=22.7(8)$ \cite{Anisovich:96}, $Q_{\mathrm{GL}}=22.1(7)$ \cite{Colangelo:18}, $Q_{\mathrm{GL}}=22.4(9)$ \cite{Kambor:96} and $Q_{\mathrm{GL}}=21.5(1.0)$ \cite{Moussallam:17}. One of the goals of this paper is to compute the ratio $Q_{\mathrm{GL}}$ in yet another independent way. 

There is a somewhat more phenomenological approach to discussing the quark mass problem in the literature, suggested by Kaplan and Manohar \cite{Manohar:86}. It is based on the setting bounds at the second-order contributions to the pseu\-doscalar-meson masses, rather than bounding the individual unmeasurable coefficients of the second-order operators. In this case, the allowed values of $m_u/m_d$ and $m_s/m_d$ are consistent with phenomenological values of pseudoscalar-meson masses to second order in quark masses and, as a consequence, belong to some curve that is certainly not an ellipse (due to approximations made in \cite{Manohar:86}, this is a second-order curve reducible to a parabola). 

Of particular interest in the studies \cite{GL:85} and \cite{Manohar:86} is the form of the implicit function $f(x,y)=0$ relating the current-quark-mass ratios $x=m_u/m_d$, $y=m_s/m_d$. This is partly due to the fact that the combined use of this theoretical information with the results of lattice simulations, which have now reached high accuracy \cite{FLAG:22}, allows one to significantly reduce the error bars when extracting quark mass ratios. But how reliable are our theoretical ideas about the function $f(x,y)=0$? 

The results of \cite{Colangelo:18} indicate that the low-energy theorem, i.e, Eq.\,(\ref{GL}), works well. This means that the higher order contributions amount to remarkably small corrections to the current-algebra estimates. However, if the current-algebra result does really play a dominant role at the physical values of $m_u$, $m_d$ and $m_s$, then the results of the above mentioned approaches should coincide (up to computational errors). This is what we want to verify in this paper, and this is the second aim of our research.

Going beyond the current-algebra results requires a careful consideration of the electromagnetic contributions to the self-energy of charged pseudoscalars. At this stage the Dashen theorem is violated, and, as calculations have shown, this essentially affects the value of the quark mass ratio $m_u/m_d$. This was noticed both on the lattice ($m_u/m_d =0.512(6)$ \cite{Duncan:96}) and in the two-loop calculations in $\chi$PT ($m_u/m_d=0.46(9)$ \cite{Amoros:01}). The question has a rich history \cite{Donoghue:92, Bijnens:93, Donoghue:93, Urech:94, Bijnens:97, Moussallam:02}. Recently, FLAG averaged the $N_f=2+1+1$ results of RM123 \cite{RM:17} and MILC \cite{MILC:18} lattice collaborations with the value of $(\Delta m_K^2)^\gamma$ from BMW \cite{BMW:15} to estimate the parameter $\epsilon$ related to the violations of Dashen's theorem  
\begin{equation}
\epsilon = \frac{(\Delta m_K^2 - \Delta m_\pi^2)^\gamma}{\Delta m_\pi^2}\simeq \frac{(\Delta m_K^2)^\gamma}{\Delta m_\pi^2}-1,
\end{equation}
where $\Delta m_K^2=m^2_{K^+}-m^2_{K^0}$ and $\Delta m_\pi^2=m^2_{\pi^+}-m^2_{\pi^0}$. The last step in this formula corresponds to the leading-order in the isospin-breaking expansion. The superscript $\gamma$ denotes corrections that arise from electromagnetic effects only. The obtained estimate $\epsilon =0.79(6)$ significantly improved the result based on the low-energy theorem $\epsilon =0.70(28)$ \cite{FLAG:17}. In the following, to take into account the violation of Dashen's theorem, we will use the latest result of FLAG. This step significantly improves the accuracy of the calculations and, together with the existing estimate of $S=27.30(34)$, allows us to obtain a more accurate result for the ratio $m_u/m_d$ which agrees with the lattice data listed in Table 9 of Ref.\,\cite{FLAG:22}. This is a third goal of our study.  

The material of the paper is distributed as follows. In Sec.\,\ref{sec2} we discuss the basic assumptions of the approach used. Then we show that the function $f(x,y)$, in the case of bounding the second order contributions to the meson masses, is a cubic curve with a number of remarkable properties. One such property is the parameterization by means of the Weinberg quantities $\bar r_x$ and $\bar r_y$, which, unlike the ellipse case, act as two independent variables. In Sec.\,\ref{sec3}, based on existing constraints on the values of $S$ and $\epsilon$, we find physical regions corresponding to admissible values of $m_u/m_d$ and $m_s/m_d$ both for the case of an ellipse and a cubic curve. This comparison allows us to conclude that the approaches are self consistent. In Sec. \ref{sec4}, the absolute values of quark masses are calculated. The additional freedom associated with the cubic curve allows us to take this step. In Sec.\,\ref{sec5}, we show that considering higher order corrections does not affect the cubic curve. This important property explains some approximations made when calculating the individual quark masses. The results are summarized in Sec.\,\ref{sec6}.

\section{Cubic curve vs ellipse}
\label{sec2}
The calculation of corrections to the current-algebra results is usually based on the effective chiral Lagrangian. Such a Lagrangian was, for example, proposed in \cite{Leutwyler:96a,Leutwyler:96b} and represents the $1/N_c$ expansion of chiral perturbation theory \cite{GL:85}. The inclusion of an additional parameter ($N_c$ is the number of colors) has several reasons. One is the consistent description of the axial $U(1)$ anomaly. The other is the wish to eliminate from the theory the unphysical symmetry of the chiral Lagrangian with respect to the Kaplan-Manohar transformations \cite{Manohar:86}. The latter is a source of an ambiguity that affects phenomenological determinations of the quark mass ratios. The large-$N_c$ limit also represents the only coherent theoretical explanation of the Okubo-Zweig-Iizuka rule \cite{Witten:79}. 

Here our study is based on the $1/N_c$ Nambu -- Jona-Lasinio (NJL) model \cite{Osipov:23a}. The effective Lagrangian of the model is obtained by calculating the one-loop quark diagrams if one additionally accepts the counting rules of large-$N_c$ chiral perturbation theory: $1/N_c = O(\delta )$, $p^2 = O(\delta )$, $m_q = O( \delta )$ \cite{Leutwyler:96a,Leutwyler:96b}. The Fock-Schwinger proper time method allows one to carry out calculations directly in coordinate space and consistently take into account the effects of explicit violation of chiral symmetry \cite{Osipov:21a,Osipov:21b,Osipov:21c}. We emphasize that for the issues discussed below there are no fundamental differences between these two approaches. Moreover, given that the relation between the NJL model and $\chi$PT is well known \cite{Arriola:91,Bruno:93,Osipov:96}, the computational results below are equally applicable to either version of the model. The only requirement is a correct account of isospin symmetry breaking, which, as is known, leads to the following mass formulas for the pions and kaons in the NLO approximation:
\begin{equation}
\label{mesonmasses}
\bar m_{ij}^2=\bar\mu_{ij}^2[1+\Delta (m_i+m_j)] +\mathcal O(\delta^3),
\end{equation}
where the low-energy constant $\Delta$ is model dependent. For instance, within the framework of large-$N_c$ chiral perturbation theory $\Delta=8B_0(2L_8-L_5)/F^2$ \cite{Leutwyler:96a,Leutwyler:96b}, in the large-$N_c$ NJL model $\Delta=\delta_M/2M_0$ \cite{Osipov:23a}. The value of $\Delta$ does not affect the form of the function $f(x,y)$, so it is not essential at this stage to specify the values of the low-energy constants entering $\Delta$. 
In the following, to fix $\Delta$ we will use the known lattice estimates. Along with formula (\ref{mesonmasses}), this is one of our main assumptions. It additionally reduces the model dependence of the result. Note also that chiral logarithms generated by one-loop meson graphs are of order $\mathcal O(\delta^3)$, and therefore appear only at the next order of the large-$N_c$ expansion.     

Counting parameters in Eq.\,(\ref{mesonmasses}) reveals that there are four degrees of freedom (e.g., $m_dB_0$, $m_d\Delta$, $x$, $y$) for fitting three masses of $\pi^+$, $K^+$ and $K^0$ mesons. This explains why $m_s/m_d$ may be expressed as a function of $m_u/m_d$ and the physical masses $m_{\pi^+}$, $m_{K^+}$,  $m_{K^0}$. 

Indeed, proceeding from the ratios
\begin{eqnarray}
\label{ratios}
\bar r_x&=&\frac{\bar m^2_{K^+}-\bar m^2_{K^0}+\bar m^2_{\pi^+}}{\bar m^2_{K^0}-\bar m^2_{K^+}+\bar m^2_{\pi^+}}, \nonumber \\ 
\bar r_y&=&\frac{\bar m^2_{K^+}+\bar m^2_{K^0}-\bar m^2_{\pi^+}}{\bar m^2_{K^0}-\bar m^2_{K^+}+\bar m^2_{\pi^+}},
\end{eqnarray}
we arrive to an implicit equation relating $x$ and $y$ 
\begin{equation}
\label{cubic}
(y^2-1)(1-x\bar r_x)=(1-x^2)(y \bar r_y -1),
\end{equation}
which describes an algebraic cubic curve of genus $g=1$. This curve consists of two hyperbolic branches and one straight (hyperbolic-type) branch. Point $(\bar r_x,\bar r_y)$ belongs to the latter one.

The result (\ref{cubic}) does not depend on the choice of starting ratios in (\ref{ratios}). Any two independent real fractional-linear functions composed of the squares of the meson masses lead to Eq.\,(\ref{cubic}). Actually there is nothing special with that because that is the way it is supposed to be. Note also that if we use Dashen's theorem to relate QCD-values of the meson masses to the phenomenological ones, we obtain that $\bar r_x=x_W$, $\bar r_y=y_W$.

To obtain an ellipse it is necessary to start from the ratios whose chiral expansion completely excludes the NLO correction. For instance, chiral expansion of the following expressions  
\begin{eqnarray}
\label{R1}
&&\bar r_1=\frac{\bar m^2_{K^+}}{\bar m^2_{\pi^+}}\stackrel{\mathrm{NLO}}{=}\frac{m_s+m_u}{m_d+m_u}\left(1+ \Delta (m_s-m_d)\right), \nonumber \\
&&\bar r_2=\frac{\bar m^2_{K^0}-\bar m^2_{K^+}}{\bar m^2_{K^0}-\bar m^2_{\pi^+}} \nonumber \\
&&\ \ \  \stackrel{\mathrm{NLO}}{=}\frac{m_d-m_u}{m_s-m_u}\left(1+ \Delta (m_s-m_d)\right)
\end{eqnarray}  
shows that NLO corrections drop out from the ratio
\begin{equation}
\label{ellipse1}
\frac{\bar r_1}{\bar r_2}\stackrel{\mathrm{NLO}}{=} \frac{m_s^2-m_u^2}{m_d^2-m_u^2}\equiv Q_1^2\,.
\end{equation} 
In this case, we have three degrees of freedom ($m_d\Delta$, $x$, $y$) for fitting two phenomenological values $\bar r_1$, $\bar r_2$. As a result, $x$ and $y$ belong to the ellipse. The actual values of $x$ and $y$ depend on the dimensionless parameter $m_d\Delta$. 
 
 It is obvious that if any real number is subtracted from both sides of Eq.\,(\ref{ellipse1}), the curve described by the equation will not change. In particular, subtracting 1 gives   
\begin{equation}
\label{ellipse2}
\frac{\bar m^2_{K^0}}{\bar m^2_{\pi^+}} \left(\frac{\bar m^2_{K^+}-\bar m^2_{\pi^+}}{\bar m^2_{K^0}-\bar m^2_{K^+}}\right)\stackrel{\mathrm{NLO}}{=} \frac{m_s^2-m_d^2}{m_d^2-m_u^2}\equiv Q_2^2,
\end{equation} 
where $Q_2^2=Q_1^2-1$. 

\begin{figure}
\includegraphics[width=0.45\textwidth]{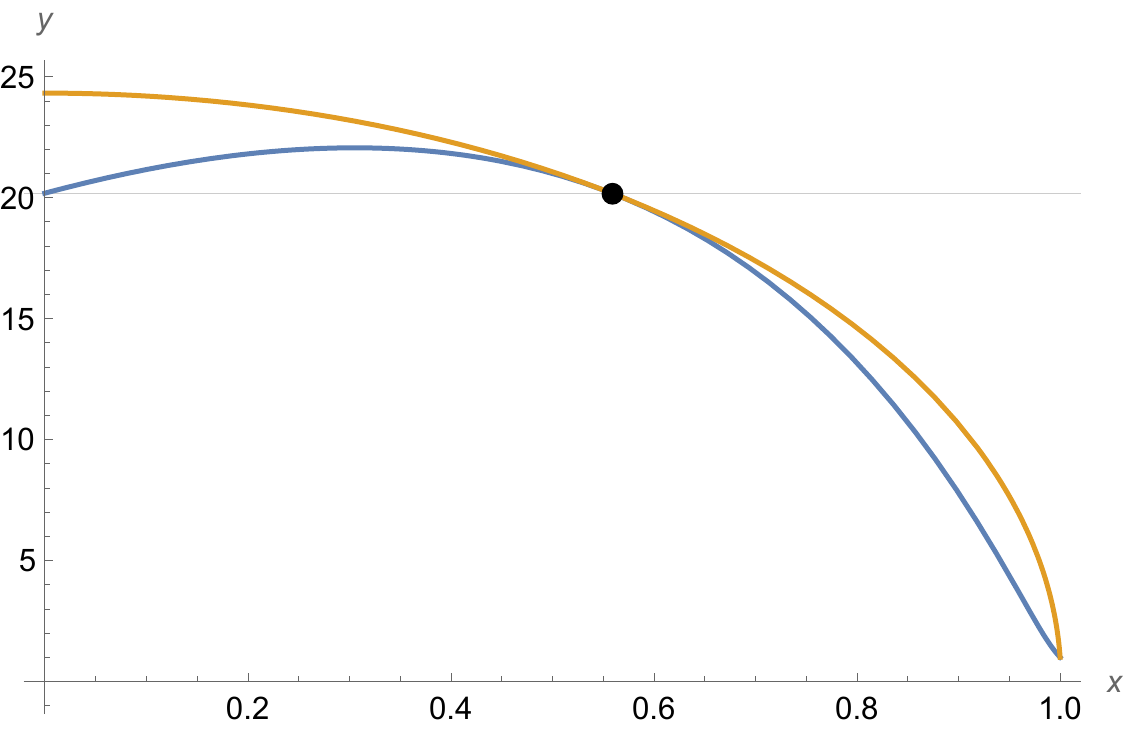}
\caption{Function $y=m_s/m_d$ vs $x=m_u/m_d$ in the NLO approximation for the case of the low-energy Gasser-Leutwyler theorem (ellipse) and for the case of the physical meson masses (cubic curve).  It is assumed that Dashen's theorem is fulfilled. The curves have a common tangent to the point $(x_W,y_W)$ and intersect at the point $(1,1)$. The horizontal line $y = y_W$ intersects the cubic curve at the points $(x_W,y_W)$ and $(0,y_W)$.}
\label{fig1}      
\end{figure}

Rewriting the left-hand side of (\ref{ellipse1}) in terms of (\ref{ratios}), we get  
\begin{equation}
\label{ellipse}
(y^2-1)(1-\bar r_x^2)=(1-x^2)(\bar r_y^2-1).
\end{equation}
Comparing Eqs.\,(\ref{cubic}) and (\ref{ellipse}) (see Fig.\,\ref{fig1}), it can be seen that both curves are parameterized by Weinberg values (\ref{Wv}). In the case of an ellipse, they form one common factor which determines its semi-major axis 
\begin{equation}
Q_1=\sqrt{Q_2^2+1}\stackrel{\mathrm{NLO}}{=}\sqrt{\frac{\bar r_y^2-\bar r_x^2}{1-\bar r_x^2}}=24.3.
\end{equation} 
In contrast, in the case of a cubic curve $\bar r_x$ and $\bar r_y$ contribute independently. As a consequence, the quark mass ratio $Q_1^2$ is not fixed along the curve. The fixed quantities here are the masses of $K^\pm$, $K^0$ and $\pi^\pm$ mesons. The cubic curve intersects the $y$-axis at the point $y=y_W$ and the $x$-axis at the point $x=x_W$ (this point belongs to another branch of the curve and is not shown in Fig.\,\ref{fig1}).

The advantage of the cubic curve is its universality. An arbitrarily chosen pair
\begin{eqnarray}
\label{genrat}
\bar r_\alpha=\frac{(\alpha_P, \bar m^2_{P})}{(\bar \alpha_P, \bar m^2_{P})}, \quad \bar r_\beta=\frac{(\beta_P, \bar m^2_{P})}{(\bar \beta_P, \bar m^2_{P})},
\end{eqnarray}
where $(\alpha_P, \bar m^2_{P})\equiv \alpha_{K^+}\bar m^2_{K^+}+\alpha_{K^0}\bar m^2_{K^0}+\alpha_{\pi^+}\bar m^2_{\pi^+}$, leads, at the next-to-leading order, to the equation 
\begin{eqnarray}
\label{arbeq}
&&(\alpha_P,\bar \mu_P^4)(\beta_P,\bar \mu_P^2)-(\alpha_P,\bar \mu_P^2)(\beta_P,\bar \mu_P^4) \nonumber \\
&+&\bar r_\alpha [(\bar \alpha_P,\bar \mu_P^2)(\beta_P,\bar \mu_P^4)-(\bar \alpha_P,\bar \mu_P^4)(\beta_P,\bar \mu_P^2)] \\
&+&\bar r_\beta [(\bar \beta_P,\bar \mu_P^4)(\alpha_P,\bar \mu_P^2)-(\bar \beta_P,\bar \mu_P^2)(\alpha_P,\bar \mu_P^4)] \nonumber \\
&=&\bar r_\alpha \bar r_\beta [(\bar \beta_P,\bar \mu_P^4)(\bar \alpha_P,\bar \mu_P^2)-(\bar \beta_P,\bar \mu_P^2)(\bar\alpha_P,\bar \mu_P^4)]. \nonumber
\end{eqnarray}
The dependence on arbitrary parameters $\alpha_P, \beta_P, \bar\alpha_P, \bar\beta_P$ can be factorized. For this purpose, $\bar r_\alpha$ and $\bar r_\beta$ need to be expressed in terms of $\bar r_x$ and $\bar r_y$. As a result, Eq.\,(\ref{arbeq}) takes the form
\begin{eqnarray}
\label{univ}
&&\mathcal F(\alpha_P, \beta_P, \bar\alpha_P, \bar\beta_P, \bar r_x,\bar r_y) \nonumber \\
&\times&[(y^2-1)(1-x\bar r_x)-(1-x^2)(y\bar r_y-1)]=0 
\end{eqnarray}
If $\mathcal F \neq 0$, then Eq.\,(\ref{cubic}) holds. Otherwise, $\bar r_\alpha$ and $\bar r_\beta$ are not independent. 

Thus, different sets in Eq.\,(\ref{genrat}) correspond to the same curve. Moreover, in the specific case $\bar\alpha_P = \bar \beta_P$ the ratios (\ref{genrat}) lead to the curve (\ref{cubic}) regardless of whether chiral expansion of the meson mass ratios is performed or not. 

Indeed, the chiral expansion allows us to pass from $\bar r_\alpha$ to an expression equivalent to (\ref{R1}) 
\begin{equation}
\bar r_\alpha =k_\alpha [1+l_\alpha m_d \Delta +{\mathcal O}(\delta^2)],
\end{equation}  
where coefficients $k_\alpha$ and $l_\alpha$ are functions of quark mass ratios $x$, $y$, and parameters $\alpha_P$, $\bar \alpha_P$
\begin{eqnarray}
k_\alpha &=&\frac{(\alpha_P, \bar\mu^2_P)}{(\bar\alpha_P, \bar\mu^2_P)}, \nonumber \\
l_\alpha &=&\frac{1}{B_0m_d}\left[\frac{(\alpha_P, \bar\mu^4_P)}{(\alpha_P, \bar\mu^2_P)}-\frac{(\bar\alpha_P, \bar\mu^4_P)}{(\bar\alpha_P, \bar\mu^2_P)}\right].
\end{eqnarray} 
Obviously, if we consider two elements of this set, $\bar r_\alpha$ and $\bar r_\beta$, we can exclude the dependence on $m_d \Delta$ and find the equation relating $y$ and $x$
\begin{equation}
\label{expeq}
k_\alpha k_\beta (l_\alpha -l_\beta)=k_\alpha l_\alpha \bar r_\beta -k_\beta l_\beta \bar r_\alpha. 
\end{equation}
It is easy to see that in the special case, $l_\alpha =l_\beta$, we arrive to the low-energy theorem (\ref{ellipse1}). If $\bar\alpha_P=\bar\beta_P$, Eq.\,(\ref{expeq}) describes the cubic curve (\ref{cubic}). For $\bar\alpha_P \neq \bar\beta_P$, different pairs of $\bar r_\alpha, \bar r_\beta$ lead to different algebraic curves. Some consequences of this behavior have been studied in \cite{Osipov:23a}. 

It should be emphasized that the low energy theorem plays a special role in selecting a specific curve from an infinite set of possible ones. At the same time, the pair $\bar r_1$, $\bar r_2$ (as well as any other pair $\bar r_\alpha$, $\bar r_\beta$), when following the phenomenological masses of pseudo-Goldstone bosons, always leads to a cubic curve (\ref{cubic}).

\section{Violations of Dashen's theorem and quark mass ratios}
\label{sec3}
At the NLO the ratios (\ref{ratios}) should be adjusted to account for deviations from the Dashen theorem. 

The difference between the masses of charged and neutral pions is almost entirely determined by the electromagnetic self-energy contribution $\Delta^2_{\mathrm{em}}$ to the mass of the charged pion
\begin{equation}
\label{piondiff}
m^2_{\pi^+}-m^2_{\pi^0}=\bar m^2_{\pi^+}-\bar m^2_{\pi^0}+\Delta^2_{\mathrm{em}}.
\end{equation}
Here the value of $\bar m^2_{\pi^+}-\bar m^2_{\pi^0}$ is proportional to $(m_d-m_u )^2$ and has been estimated as  $\bar m_{\pi^+}-\bar m_{\pi^0}=0.17(3)\,\mbox{MeV}$ \cite{GL:85}. Together with the observed mass difference, Eq.\,(\ref{piondiff}) implies 
\begin{equation}
\label{dg1}
\Delta^2_{\mathrm{em}}=1.21(1)\times 10^{-3}\,\mbox{GeV}^2.
\end{equation}

In the case of the kaons, corrections from higher orders of the chiral expansion can be characterized with the dimensionless parameter $\epsilon$, which is defined by  \cite{FLAG:17}
\begin{equation}
\tilde\Delta^2_{\mathrm{em}}=\Delta^2_{\mathrm{em}}+\epsilon (m^2_{\pi^+}-m^2_{\pi^0}).   
\end{equation}
Once known, $\epsilon$ allows one to consistently subtract the electromagnetic part of the kaon-mass splitting to obtain the QCD splitting   
\begin{equation}
\label{kaondiff}
\bar m^2_{K^+}-\bar m^2_{K^0}=m^2_{K^+}-m^2_{K^0}-\tilde\Delta^2_{\mathrm{em}}.
\end{equation}

In a recent review of results obtained on the lattice by various collaborations, FLAG \cite{FLAG:22} gives the following averaged values for $\epsilon$
\begin{eqnarray}
\label{Nf4}
\epsilon &=& 0.79(6) \quad (N_f=2+1+1), \\
\label{Nf3}
\epsilon &=& 0.73(17) \quad (N_f=2+1), 
\end{eqnarray}
which are calculated with four $N_f=4$ and  three $N_f=3$ quark flavors. Based on (\ref{Nf4}) we get  
\begin{equation}
\label{dg2}
\tilde\Delta^2_{\mathrm{em}}=2.21(8)\times 10^{-3}\,\mbox{GeV}^2.
\end{equation}

Numerical estimates of $\Delta^2_{\mathrm{em}}$ and $\tilde\Delta^2_{\mathrm{em}}$ allow us to establish bounds on the values of the parameters (\ref{cubic}) and (\ref{ellipse}) in the case where Dashen theorem is violated
\begin{eqnarray}
r_x&=&\frac{m^2_{K^+}\!-\!m^2_{K^0}\!+\!m^2_{\pi^+}\!-\!\tilde\Delta^2_{\mathrm{em}}\!-\!\Delta^2_{\mathrm{em}}}{m^2_{K^0}\!-\!m^2_{K^+}\!+\!m^2_{\pi^+}\!+\!\tilde\Delta^2_{\mathrm{em}}\!-\!\Delta^2_{\mathrm{em}}}=0.498(5), \nonumber \\ 
r_y&=&\frac{m^2_{K^+}\!+\!m^2_{K^0}\!-\!m^2_{\pi^+}\!-\!\tilde\Delta^2_{\mathrm{em}}+\Delta^2_{\mathrm{em}}}{m^2_{K^0}\!-\!m^2_{K^+}\!+\!m^2_{\pi^+}\!+\!\tilde\Delta^2_{\mathrm{em}}\!-\!\Delta^2_{\mathrm{em}}}=19.32(6). \nonumber
\end{eqnarray} 

In particular, using the low-energy theorem (\ref{ellipse1}), we find 
\begin{equation}
Q_1^2=\frac{(m^2_{K^+} \!-\!\tilde\Delta^2_{\mathrm{em}})(m^2_{K^0}\!-\!m^2_{\pi^+}\!+\!\Delta^2_{\mathrm{em}})}{(m^2_{\pi^+}\!-\!\Delta^2_{\mathrm{em}})(m^2_{K^0}\!-\!m^2_{K^+}\!+\!\tilde\Delta^2_{\mathrm{em}})}.
\end{equation}
This gives the following value for the semi-major axis of the ellipse $Q_1=22.28(15)$, which is in excellent agreement with the result of \cite{Colangelo:17}, $Q_{\mathrm{GL}}=22.1(7)$, and the FLAG estimate $Q_{\mathrm{GL}}=22.5(5)$ \cite{FLAG:22}. Considering the cubic curve, we find in turn $Q_{\mathrm{GL}}=22.23(16)$. Since the calculation error in (\ref{Nf4}) is rather small, the theoretical estimate of $Q_{\mathrm{GL}}$ also has a fairly high accuracy.

The quark mass ratio $S=m_s/m_{ud}=27.23(10)$ \cite{FLAG:22} is well known from lattice simulations of QCD and can therefore be used to estimate the value of $m_u/m_d$. It is worth noting that the uncertainties in the corrections to the Dashen theorem barely affect this ratio (the electromagnetic effects in this quantity are estimated to be $\simeq 0.18$\% \cite{Bazavov:18}). Thus, the lattice result for $S$ is a good approximation that can be trusted. The idea was realized in \cite{Colangelo:17}, where the authors found that $m_u/m_d=0.44(3)$, which corresponds to $Q_{\mathrm{GL}}=22.1(7)$. This estimate can now be improved due to the high precision of (\ref{Nf4}).

\begin{figure}
\includegraphics[width=0.45\textwidth]{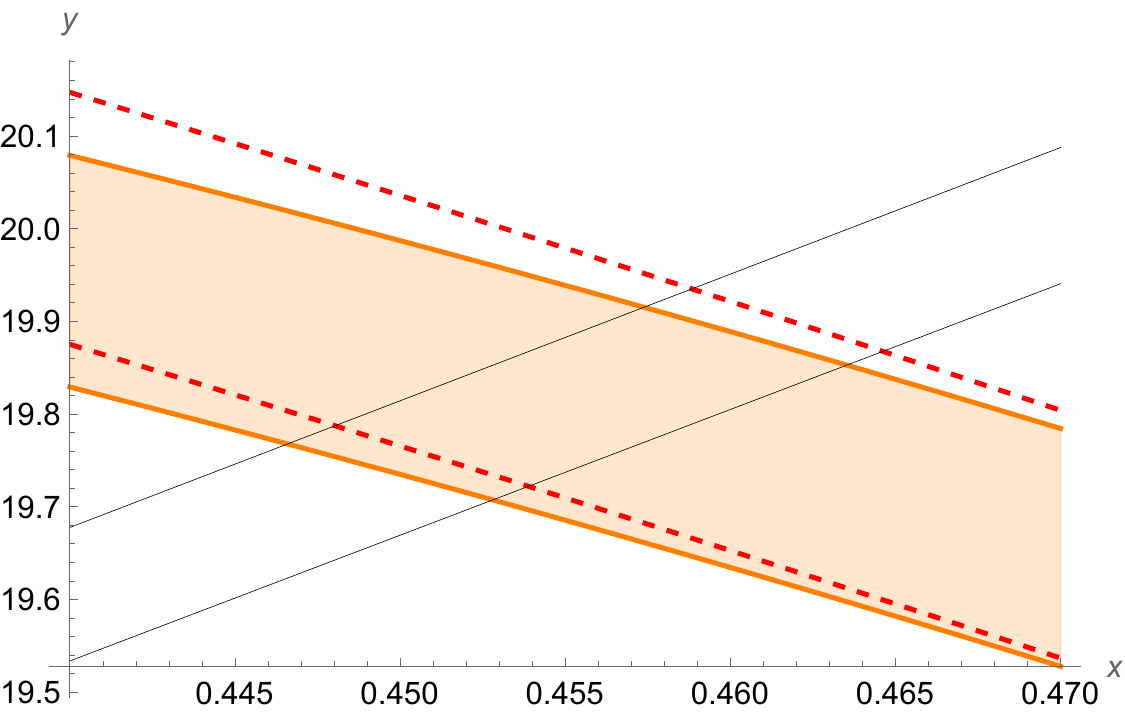}
\caption{The acceptable range for quark mass ratios in the case of an ellipse (corresponding interval is indicated by dotted lines) and a cubic curve (filled band). Thin lines correspond to the ratio $m_s/m_{ud}=27.23(10)$ given by FLAG for lattice simulations with four quark flavors \cite{FLAG:22}. Their intersection with the ellipse and the cubic curve determines the boundaries of the region of admissible values of the quark mass ratios $x=m_u/m_d$ and $y=m_s/m_d$ in each of the two cases considered.}
\label{fig2}      
\end{figure}

Fig.\,\ref{fig2} shows the result of such calculations. It can be seen that, in the considered range for $S$, calculations based on the low-energy theorem and the cubic curve are in remarkable agreement with each other. Indeed, here we have 
\begin{eqnarray}
\label{utod}
&&\left. 
\begin{array}{rcl}
    m_u/m_d&=&0.455(8)  \\
    m_s/m_d&=&19.81(10) \\
\end{array} 
\right\} \mbox{(cubic curve)}, \nonumber \\
&&
\left. 
\begin{array}{rcl}
    m_u/m_d&=&0.456(8)  \\
    m_s/m_d&=&19.83(11) \\
\end{array} 
\right\} \mbox{(ellipse)}.
\end{eqnarray}
Let us also indicate the value of the quark mass ratio $R$, which corresponds to the estimates made above: $R=35.0(6)$ (cubic curve) and $R=35.1(6)$ (ellipse).

To be precise, it should be noted that the ratio $m_u/m_d$ depends on the scale in QCD plus QED, since up and down quarks have different electric charges. However, this effect appears only in the second order of isospin breaking, which is beyond the accuracy of our calculations. The result (\ref{utod}) is compatible with the FLAG average $m_u/m_d = 0.465(24)$  ($N_f=2+1+1$) \cite{FLAG:22}, but our precision is $2.5$ times higher.
 
\begin{table*}
\label{tab1}
\caption{The light quark masses $m_u$, $m_d$, $m_s$, the isospin-averaged up- and down-quark mass $m_{ud}$, cubic root $\Sigma^{1/3}=|\langle\bar qq\rangle^{1/3}| $ of the $SU(3)$ quark condensate $\langle\bar qq\rangle$ (all in [MeV]), and quark mass ratios $m_u/m_d$, $Q_{\mathrm{GL}}$, $R$. The low-energy constant $B_0$ is given in GeV. All values refer to the $\overline{MS}$-scheme at scale $2\,\mbox{GeV}$. } 
\begin{footnotesize}
\begin{tabular*}{\textwidth}{@{\extracolsep{\fill}}llcccccccc@{}}
\hline
\hline 
\multicolumn{1}{l}{Source}
& \multicolumn{1}{c}{Details}
& \multicolumn{1}{c}{$m_u$}
& \multicolumn{1}{c}{$m_d$}
& \multicolumn{1}{c}{$m_{ud}$}
& \multicolumn{1}{c}{$m_s$}
& \multicolumn{1}{c}{$m_u/m_d$}
& \multicolumn{1}{c}{$Q_{\mathrm{GL}}$}
& \multicolumn{1}{c}{$R$}
& \multicolumn{1}{c}{$\Sigma^{1/3}$}
\\
\hline
\\[-9pt] 
This work  
& $B_0\!=\!2.682(53)$ 
& $2.14\pm 0.07$
& $4.70\pm 0.12$
& $3.420(70)$
& $93.13\pm 2.25$
& $0.455(8)$
& $22.23(16)$ 
& $35.02(61)$
& $280.71(1.86)$
\\

& $B_0\!=\!2.66$
& $2.16\pm 0.03$
& $4.74\pm 0.03$
& $3.447(2)$
& $93.85\pm 0.41$
& $0.455(8)$
& $22.23(16)$ 
& $35.02(61)$
& $280$
\\

& $B_0\!=\!2.516(67)$ 
& $2.28\pm 0.09$
& $5.01\pm 0.16$
& $3.647(100)$
& $99.31\pm 3.08$
& $0.455(8)$
& $22.23(16)$ 
& $35.02(61)$
& $274.79(2.45)$
\\

PDG \cite{PDG:22}
& $\ \ \ \ \ \ \ -$
& $2.16^{+0.49}_{-0.26}$
& $4.67^{+0.48}_{-0.17}$
& $3.45^{+0.35}_{-0.15}$
& $93.4^{+8.6}_{-3.4}$
& $0.474^{+0.056}_{-0.074}$
& $-$
& $-$ 
& $-$
\\
FLAG \cite{FLAG:22}  
& $N_f\!=\!2\!+\!1\!+\!1$
& $2.14\pm 0.08$
& $4.70\pm 0.05$
& $3.410(43)$
& $93.44\pm 0.68$ 
& $0.465(24)$
& 22.5(5)
& $35.9(1.7)$
& $286\pm 23$
\\
   
& $N_f\!=\!2\!+\!1\!$
& $2.27\pm 0.09$
& $4.67\pm 0.09$
& $3.364(41)$
& $92.03\pm 0.88$ 
& $0.485(19)$
& 23.3(5)
& $38.1(1.5)$
& $272 \pm 5$
\\

\hline
\hline
\end{tabular*}
\end{footnotesize} 
\end{table*}

  
\section{Absolute values of quark masses }
\label{sec4}
It is well known that the running quark masses $m_q (\mu)$ in QCD depend on the renormalization-group scale $\mu$ in the $\overline{MS}$ subtraction scheme. The same can be said about the magnitude of the quark condensate $\langle \bar qq\rangle$, associated with the low-energy constant $B_0$. However, their product $m_q B_0$ is a renormalization-group invariant quantity. Since this product is one of the intrinsic parameters on the cubic curve, we can estimate the masses of individual quarks if $B_0$ is known, and vice versa. The lattice calculations of $B_0(\mu)$ are currently not highly accurate. For our estimates here we will use the NLO result $B_0(2\,\mbox{GeV}) =2.682(36)(39)\,\mbox{GeV}$ \cite{Borsanyi:13}, the total error of which is relatively small $\sim 2$\%. This value is reasonably consistent with the high-quality entries in Table 21 of Ref.\,\cite{FLAG:22}.  

To determine the absolute values of quark masses, we solve three equations (\ref{mesonmasses}) for a given value of $B_0$. The parameter $\Delta$ is varied so that the solution belongs to the region of intersection of the cubic curve with the interval $S=27.23(10)$ (see Fig.\,\ref{fig2}). The corresponding results are presented in Table\,I. All results are given in the $\overline{MS}$ scheme at $2\,\mbox{GeV}$, which is standard nowadays. 

The first line in Table\,I contains our results corresponding to the values of $B_0=2.682(53)\,\mbox{GeV}$. To get an idea of the role of errors in the value of $B_0$, we also present the results of similar calculations performed at a strictly fixed value of $B_0=2.66\,\mbox{GeV}$ (see second row in Table\,I). For that we took the arithmetic mean of the central values of \cite{Borsanyi:13} and ETM collaboration \cite{ETM:10}. In the third row, we also present results corresponding to the value of $B_0(2\,\mbox{GeV})=2.516(67)\,\mbox{GeV}$ \cite{ETM:21}, which is obtained by ETM collaboration with the use of $N_f=2+1+1$ flavors of twisted mass fermions by matching to the LO of $\chi$PT. This allows us to trace how a change in the accuracy of the $\chi$PT calculations affects the individual quark masses. The fourth line shows the values quoted by the Particle Data Group \cite{PDG:22}, and the fifth and sixth lines represent the FLAG estimates \cite{FLAG:22} for the case of four and three quark flavors correspondingly. They are obtained by averaging the results of the different lattice collaborations in accord with the FLAG's selection criteria.

A direct comparison of our estimates presented in the first three rows of Table.\,I shows that an error of $\simeq 2\%$ in the value of $B_0$, together with the already existing $\simeq 1.5\%$ uncertainty in the determination of the values of $\epsilon$ and $S$, leads to an $\simeq 3\%$ error in the quark masses. This is enough to claim that our results are in full agreement with the data presented by the PDG, as well as the average values quoted by FLAG. This agreement allows us to conclude that it is reasonable to use the $\delta$-expansion, which is behind the original formula of our analysis (\ref{mesonmasses}). In the next section, we will return to discuss the results listed in the first and third rows of Table\,I, 
but for that we must first clarify the role of higher order corrections in the approach considered.

\section{Higher order $1/N_c$ corrections}
\label{sec5}
Let us turn to the study of the role played by higher order corrections in the case of a cubic curve. Consider ratios (\ref{genrat}), where now we expect that the mass squared $\bar m^2_P$ have next-to-next-to-leading order (NNLO)  contributions $\sim {\mathrm O}(\delta^3)$
\begin{equation}
\label{mP2}
\bar m^2_P \to Z_P \bar M^2_P =\bar m^2_P + \delta\bar m^2_P. 
\end{equation}  
Here $\bar m^2_P$ is given by Eq.(\ref{mesonmasses}), $\delta\bar m^2_P={\mathrm O(\delta^3)}$ is the NNLO part, and $Z_P$ is due to renormalization of the one-loop diagrams. The explicit expressions for $\delta\bar m^2_P$ in $1/N_c\chi$PT are well known (see, for example, \cite{Oller:15} and references therein), so we will only discuss general issues concerning their role with respect to the results obtained above.  

Eq.(\ref{mP2}) can be written as 
\begin{eqnarray}
\label{nnlo}
\bar M^2_{ij}&=&B_0(m_i+m_j)[1+\Delta (m_i+m_j) +c_{ij}] \nonumber \\
&=&\bar m^2_{ij}(1+c_{ij}), 
\end{eqnarray}
where $c_{ij}=\{c_{ud}\equiv c_{\pi^+}, c_{us}\equiv c_{K^+}, c_{ds}\equiv c_{K^0}\}={\mathrm O(m_q^2)}$ already contains a renormalization of the one-loop diagrams at the expense of $Z_P$. The last equality in (\ref{nnlo}) is satisfied up to and including the NNLO terms.

The first thing we must realize is that NNLO contribution has no effect on the shape of the cubic curve. Indeed, the ratios (\ref{genrat}) is now constructed from the physical masses $\bar M^2_P$, but the result
\begin{equation}
\bar R_\alpha =\frac{(\alpha_P , \bar M^2_P)}{(\bar\alpha_P ,\bar M^2_P)}=\frac{(\alpha'_P , \bar m^2_P)}{(\bar\alpha'_P, \bar m^2_P)}
\end{equation}
 differs from (\ref{genrat}) only by the values of the coefficients $\alpha'_P =(1+c_P)\alpha_P$  and $\bar\alpha'_P =(1+c_P)\bar\alpha_P$. As we already learn, the magnitude of these coefficients does not affect the equation of the cubic curve due to universality (\ref{univ}). Thus, $x$ and $y$ still belong to the curve (\ref{cubic}) after accounting for the NNLO corrections. 
 
The same conclusion can be reached in another way. In Eq.(\ref{nnlo}), the factor $(1+c_{ij})$ is the only effect of the NNLO contributions on the meson masses. It can be absorbed by redefining the quark masses $(m_i+m_j)(1+c_{ij})=(m_i'+m_j')$, i.e., 
\begin{eqnarray}
m_u'&\!=\!&m_u\!+\!\frac{1}{2}\!\left(m_u c^{(+)}_{K^+\pi^+}\!-\!m_d c^{(-)}_{K^0\pi^+}\!+\!m_s c^{(-)}_{K^+K^0}\! \right)\!, \nonumber \\
m_d'&\!=\!&m_d\!+\!\frac{1}{2}\!\left(m_d c^{(+)}_{K^0\pi^+}\!-\!m_u c^{(-)}_{K^+\pi^+}\!-\!m_s c^{(-)}_{K^+K^0} \! \right)\!, \\
m_s'&\!=\!&m_s\!+\!\frac{1}{2}\!\left(m_s c^{(+)}_{K^+K^0}\!+\!m_u c^{(-)}_{K^+\pi^+}\!+\!m_d c^{(-)}_{K^0\pi^+} \!\right)\!,\nonumber 
\end{eqnarray}
where $c^{(\pm)}_{PP'}=c_P\pm c_{P'}$. As a result, Eq.(\ref{nnlo}) retains the form of Eq.(\ref{mesonmasses})
\begin{equation}
\label{nnlo2}
\bar M^2_{ij}=B_0(m_i'+m_j')[1+\Delta (m_i'+m_j')]. 
\end{equation}
This means that if we repeat the calculations of Sec.\,\ref{sec2} using (\ref{nnlo2}), we again arrive at Eq.\,(\ref{cubic}).  

The individual quark masses obtained with the formulae (\ref{nnlo2}) and the physical region of Fig.\,\ref{fig2} depend on the particular value of $B_0$. In Table I, we consider two different values taken from papers \cite{Borsanyi:13} and \cite{ETM:21}. 

In \cite{Borsanyi:13}, $B_0$ results from the NLO-based fit to the SU(2) $\chi$PT (the authors also studied the effect of NNLO corrections, but unfortunately do not give the numerical value of $B_0$ for this case). The average value of the light quark masses reported by the BMW collaboration \cite{BMW:10} was used to find  $B_0=2.682(53)\,\mbox{GeV}$. Therefore, it seems reasonable to compare the values of quark masses listed in the first row of Table I  with the estimates of BMW collaboration: $m_u=2.15(03)(10)\,\mbox{MeV}$, $m_d=4.79(07)(12)\,\mbox{MeV}$, $m_{ud}=3.469(47)(48)\,\mbox{MeV}$ and $m_s=95.5(1.1)(1.2)\,\mbox{MeV}$ \cite{BMW:10}. Obviously, there is complete agreement of the data here.
 
\begin{table*}
\label{tab2}
\caption{Corrections to the current algebra results for the quark mass ratios $Q_{\mathrm{GL}}$, $S$ and $R$.} 
\begin{footnotesize}
\begin{tabular*}{\textwidth}{@{\extracolsep{\fill}}lllllll@{}}
\hline
\hline 
\multicolumn{1}{l}{Source}
& \multicolumn{1}{c}{$Q_{\mathrm{GL}}$}
& \multicolumn{1}{c}{$S$}
& \multicolumn{1}{c}{$R$}
& \multicolumn{1}{c}{$\Delta_S$}
& \multicolumn{1}{c}{$\Delta_R$}
& \multicolumn{1}{c}{$\Delta_Q$}
\\
\hline
\\[-9pt] 
BMW  \cite{Fodor:16} 
& $23.40(64)$ 
& $27.53(22)$ 
& $38.20(1.95)$
& $-0.063$
& $-0.028$
& $-0.089$
\\

RM123  \cite{Giusti:17}
& $23.8(1.1)$ 
& $-$
& $40.4(3.3)$
& $-0.042$
& $-0.060$
& $-0.099$
\\

CLLP  \cite{Colangelo:18} 
& $22.1(7)$ 
& $27.30(34)$
& $34.2(2.2) $
& $-0.051(12)$
& $0.053(14)$
& $0$
\\

Leutwyler  \cite{Leutwyler:23} 
& $22.4(2)$ 
& $27.23(10)$
& $35.5(5)$
& $-0.048(5)$
& $0.037(49)$
& $-0.011(45)$
\\

This work  
& $22.23(16)$ 
& $27.23(10)$
& $35.02(61)$
& $-0.051(4)$
& $0.058(5)$
& $0.0032(5)$
\\
\hline
\hline
\end{tabular*}
\end{footnotesize} 
\end{table*}

On the contrary, in \cite{ETM:21} the fit to the LO $\chi$PT is used to find the value $B_0=2.516(67)\,\mbox{GeV}$. In this case (see the third row of Table I), one also observes a full agreement with the estimates of \cite{ETM:21}: $m_{ud}=3.636(66)(^{+60}_{-57})\,\mbox{MeV}$,  $m_s=98(7)(2.4)(^{+4}_{-3.2})\,\mbox{MeV}$. 

In both cases a powerful tandem: Eq.\,(\ref{nnlo2}) plus cubic curve, copes well with the calculation of individual quark masses.  The decisive property here is the universality of the cubic curve, the shape of which does not depend on higher order corrections. We can conclude that if $B_0$ is known with a given accuracy, then the second parameter $\Delta$ and, as a consequence, the individual quark masses are uniquely fixed by the admissible region shown in Fig.\,\ref{fig2}. 

The final topic to be considered here is whether formula (\ref{nnlo2}) is suitable for evaluating corrections to the quark mass ratios $S$, $R$ and $Q_{\mathrm{GL}}$. These corrections are widely discussed in the literature \cite{Colangelo:18, Fodor:16, Giusti:17, Leutwyler:23}, as they are important for the justification of the low-energy theorem. 

Let us recall the relation between the corresponding quantities and the ratios of the squared meson masses  
\begin{eqnarray}
\label{rR}
&& 2 \frac{\bar M_K^2}{\bar M_{\pi^+}^2} = (S+1)(1+\Delta_S), \nonumber \\
&&\frac{\bar M^2_K - \bar M^2_{\pi^+}}{\bar M^2_{K^0}-\bar M^2_{K^+}}=R (1+\Delta_R ), \nonumber \\
&&  \frac{\bar M_K^2}{\bar M_{\pi^+}^2} \frac{\bar M^2_K - \bar M^2_{\pi^+}}{\bar M^2_{K^0}-\bar M^2_{K^+}} =  Q^2_{\mathrm{GL}} (1+\Delta_Q).
\end{eqnarray}
Since $2 Q^2_{\mathrm{GL}}=R(S+1)$, the correction $\Delta_Q$ is expressed in terms of $\Delta_S$ and $\Delta_R$, that is, $\Delta_Q =\Delta_S+\Delta_R+\Delta_S\Delta_R$. To compute $\Delta_S$ and $\Delta_R$, we use Eq.\,(\ref{nnlo2}). This yields 
\begin{eqnarray}
\Delta_S&=&\frac{\Delta (m_s^2-m_um_d)}{(m_s+m_{ud})[1+2\Delta m_{ud})]}, \nonumber \\
\Delta_R&=&\frac{-\Delta (m_s-m_u)(m_s-m_d)}{(m_s-m_{ud})[1+2\Delta (m_s+m_{ud})]}.
\end{eqnarray}
Numerical estimates of these quantities, obtained using the quark masses presented in the first row of Table I, are collected in the last row of Table II.

These estimates show that the low-energy theorem is fulfilled with a high degree of accuracy. The correction $\Delta_Q=0.0032(5)$ is small, and in this conclusion our result is in agreement with calculations performed in the framework of $\chi$PT and independently in the dispersive approach \cite{Colangelo:18}. A recent update of the higher-order corrections \cite{Leutwyler:23} shows that the central value of $\Delta_Q$ is indeed small, but due to the large uncertainty it is still possible that $\Delta_Q$ has the same magnitude as $\Delta_S$. The latter is indicated by the data of the lattice calculations, which we present here following the Table 10 of \cite{Colangelo:18}. Lattice calculations should obviously clarify this situation in the future.

\section{Summary}
\label{sec6}
We have shown that the cubic curve is another useful source for obtaining information on quark masses. Indeed, the ellipse represents the region allowed by the low-energy theorem for $Q_{\mathrm GL}$, and the cubic curve represents the region allowed by the lattice result for $\epsilon$. As we established above, these regions are different, but for the interval given by the quark mass ratio $S$, this difference turns out to be negligible. As a consequence, the results of both approaches are in excellent agreement with each other. 

This fact allowed us to more seriously consider the next alternative, the cubic curve instead of the ellipse, since it allows us to estimate the individual values of the quark masses. The resulting estimates turned out to be quite reasonable and, in principle, would be more accurate if the precision of the value of the low-energy constant $B_0$ were improved.

Finally, we examined the role of higher order $1/N_c$ corrections on the presented results and showed that the universality condition ensures that the equation of the cubic curve is not affected by the NNLO corrections.

\section{Acknowledgments}
I am grateful to B. Hiller, D.\,I Kazakov, V.\,A. Osipov and O.\,V. Teryaev for their interest to this work and stimulating discussions.

\end{document}